\documentclass[english]{article}
\usepackage[T1]{fontenc}
\usepackage[latin9]{inputenc}
\usepackage{amsmath}
\usepackage{amssymb}
\usepackage{graphicx}

\makeatletter
\newcommand{\lyxaddress}[1]{
\par {\raggedright #1
\vspace{1.4em}
\noindent\par}
}

\makeatother

\usepackage{babel}
\begin{document}

\title{Isospin symmetry breaking and the neutron-proton mass difference}

\author{N.F. Nasrallah}

\maketitle

\lyxaddress{\begin{center}
Lebanese University, Faculty of Science, Tripoli, Lebanon
\par\end{center}}
\begin{abstract}
QCD sum rules using polynomial kernels are used to evaluate the strong
part of the proton-neutron mass difference $\delta M_{np}$ in a model
independent fashion. The result for the mass difference turns out
to depend sensitively on the value of the four quark condensate $\langle(\overline{q}q)^{2}\rangle$
and reproduces the experimental value of $\delta M_{np}$ for $\left\langle (\bar{q}q)^{2}\right\rangle $
$\sim$ 2$\left\langle \bar{q}q\right\rangle $$^{2}$.
\end{abstract}
\newpage{}

\section{Introduction}

The QCD sum rule method introduced by Shifman, Vainshtein and Zakharov
\cite{SVZ} has extended the applicability of QCD far beyond simple
perturbation theory. The method was adapted to the case of nucleons
by Ioffe \cite{Ioffe} and independently by Chung, Dosch, Kremer and
Schall \cite{CDKS}. These authors showed how to approach one of the
fundamental problems of hadronic physics, the calculation of the baryon
masses from the Lagrangian and the vacuum condensates of QCD.

Several authors \cite{Jin et al} undertook the study of the neutron-proton
mass difference $\delta M_{np}$ using the external field method of
Ioffe and Smilga\cite{Ioffe}. These authors use Borel (Laplace) transforms
of correlators in order to suppress the unknown contributions of the
continua which appear in dispersion integrals.

I present here a different approach which uses instead as kernels
in the dispersion integrals simple polynomials \cite{kremer} the
coefficients of which are determined by the general shape of the physical
spectrum and which, as will be shown, offer stronger damping of the
unknown parts of the spectrum in addititon to better stability.

\section{The Calculation}

It has been known since the early days of QCD that the mass splitting
of hadrons in an isospin multiplet arises from two sources: virtual
photon exchange, which can be related to virtual photon scattering
by the Cottingham formula \cite{Cottingham} and isospin symmetry
breaking which arises from the difference between the masses of the
light quarks $m_{u}$and $m_{d}$ as well as the difference between
the light quark condensates $\langle\overline{u}u\rangle$ and $\langle\overline{d}d\rangle.$

Let us start by a reminder of how electromagnetic and strong parts
of the mass splittings arise \cite{n and s}

The quark mass term in the QCD Hamiltonian density reads

\begin{equation}
H(x)=m_{u}\overline{u}u+m_{d}\overline{d}d+...\label{eq:1}
\end{equation}

The effect of turning on electromagnetism is described by the additional
term

\begin{equation}
H_{em}(x)=\frac{e^{2}}{2(2\pi)^{4}}\int\frac{d^{4}q}{q^{2}}\int d^{4}xe^{iqx}Tj_{\mu}(x)j_{\mu}(0)\label{eq:2}
\end{equation}

which accounts for virtual photon exchange.

\begin{equation}
j_{\mu}=\frac{1}{3}\overline{u}\gamma_{\mu}u-\frac{2}{3}\overline{d}\gamma_{\mu}d\label{eq:3}
\end{equation}

is the electromagnetic current expressed in terms of quark fields.

Following Cottingham\cite{Cottingham} eq. \eqref{eq:1} can be transformed
into

\begin{align}
H_{em}(0) & =\frac{e^{2}}{8\pi^{2}}\int_{0}^{\infty}\frac{dQ^{2}}{Q^{2}}\int_{-Q}^{Q}dq_{0}\sqrt{Q^{2}-q_{0}^{2}}.i\int d^{4}xe^{iqx}Tj_{\mu}(x)j_{\mu}(0)\nonumber \\
 & =\frac{e^{2}}{8\pi^{2}}\int_{0}^{\lambda}\frac{dQ^{2}}{Q^{2}}\int_{-Q}^{Q}dq_{0}\sqrt{Q^{2}-q_{0}^{2}}.i\int d^{4}xe^{iqx}Tj_{\mu}(x)j_{\mu}(0)\label{eq:4}\\
 & +\frac{e^{2}}{8\pi^{2}}\int_{\lambda}^{\infty}\frac{dQ^{2}}{Q^{2}}\int_{-Q}^{Q}dq_{0}\sqrt{Q^{2}-q_{0}^{2}}.i\int d^{4}xe^{iqx}Tj_{\mu}(x)j_{\mu}(0)\nonumber 
\end{align}

where $Q^{2}=-q^{2}$ and $\lambda$ is a divider between small and
large values of $Q^{2}.$

Matrix elements of the low energy contribution to eq. \eqref{eq:4}
can be obtained from experimental observation, the insertion of intermediate
states, etc., this term we call $H_{coul.}^{\lambda}(0)$, the \textquotedbl{}Coulomb\textquotedbl{}
part. In the (divergent) high energy part we can use the operator
product expansion (OPE) of the electromagnetic currents with the result

\begin{equation}
H_{em}(0)=H_{coul}^{\lambda}(0)+\frac{e^{2}}{6\pi^{2}}\ln\frac{\Lambda^{2}}{\lambda^{2}}m_{d}\overline{d}d-\frac{e^{2}}{24\pi^{2}}\ln\frac{\Lambda^{2}}{\lambda^{2}}m_{u}\overline{u}u+...\label{eq:5}
\end{equation}

with $\Lambda$ an ultraviolet cutoff.

When this is added to expression \eqref{eq:1} the divergent terms
are absorbed in the renormalized masses and the Hamiltonian density
takes the form

\begin{equation}
H(0)=H_{coul}(0)+\frac{(m_{u}+m_{d})}{2}(\overline{u}u+\overline{d}d)+\frac{(m_{u}-m_{d})}{2}(\overline{u}u-\overline{d}d)\label{eq:6}
\end{equation}

The first and last terms in the equation above break isospin symmetry
and contribute to the neutron-proton mass difference $\delta M_{np}.$

\begin{equation}
\delta M_{np}=\langle p\left\vert H_{coul}(0)\right\vert p\rangle+\frac{\delta m}{2M_{p}}\langle p\left\vert u_{3}\right\vert p\rangle=\delta M_{coul}+\delta M_{q}\label{eq:7}
\end{equation}

with 
\[
\delta m=m_{d}-m_{u}
\]
 
\begin{equation}
u_{3}=\overline{u}u-\overline{d}d\label{eq:8}
\end{equation}

This separation depends of course on the value of the divider $\lambda$
but this dependence is extremely weak: the relative change in the
value of $\delta M_{q}$ for two values $\lambda_{1}$ and $\lambda_{2}$
is

\begin{equation}
\backsim1.5.10^{-3}\ln\frac{\lambda_{2}^{2}}{\lambda_{1}^{2}}\label{eq:9}
\end{equation}

$\delta M_{coul}$ has been estimated a long time ago \cite{g and l},
$\delta M_{coul}=-(.76\pm.30MeV)$. A recent evaluation \cite{wl}
which is here adopted is

\begin{equation}
\delta M_{coul}=(-1.30\pm.50)MeV\label{eq:10}
\end{equation}

The measured value $\delta M_{np}$ $=1.29MeV$ yields then

\begin{equation}
(\delta M_{q})_{\exp}=(2.60\pm.50)MeV\label{eq:11}
\end{equation}

The theoretical expression is given by eq. \eqref{eq:7}

\begin{equation}
\delta M_{q}=\delta mU\label{eq:12}
\end{equation}

with 
\begin{equation}
U=\langle p\left\vert u_{3}\right\vert p\rangle/2M_{p}\label{eq:13}
\end{equation}

The aim of the present calculation is to evaluate $U$ using the asymptotic
forms of the various matrix elements given by QCD and the incomplete
available knowledge about the nucleon continuum with an emphasis on
the stability of the calculation. For this purpose I follow the external
field approach of Ioffe and Smilga \cite{Ioffe} in which the quarks
are coupled to a weak external scalar field $S(x)$ through an additional
term to the QCD Lagrangian

\begin{equation}
\Delta L=-S(x)(\overline{u}u-\overline{d}d)\label{eq:14}
\end{equation}

$S(x)$ can be taken a constant. The correlation function of the nucleon
currents in the presence of $S(x)$ is

\begin{equation}
\Pi(S,q)=i\int d^{4}xe^{iq.x}\langle0\left\vert T\eta_{p}(x)\overline{\eta_{p}}(0)\right\vert 0\rangle_{S}\label{eq:15}
\end{equation}

where $\eta_{p}(x)$ is the proton interpolating field of Ioffe\cite{Ioffe}

\begin{equation}
\eta_{p}(x)=\epsilon_{abc}U_{a}^{T}(x)C\gamma_{\mu}U_{b}(x)\gamma_{5}\gamma_{\mu}d_{c}(x)\label{eq:16}
\end{equation}

where a, b, c stand for colour indices and $C=-C$$^{T}$ is the charge
conjugation matrix. Lorentz covariance and parity allow the decomposition

\begin{equation}
\Pi(S,q)=\Pi^{1}(S,q)+\gamma.q\Pi^{q}(S,q)\label{eq:17}
\end{equation}

To first order in $S$ the two invariant functions can be written
as

\begin{equation}
\Pi^{i}(S,q)=\Pi_{0}^{i}(q^{2})+S\Pi_{1}^{i}(q^{2})\label{eq:18}
\end{equation}

where $\Pi_{0}^{i}$ denote the invariant functions in the absence
of the external field and $\Pi_{1}^{i}$ are the linear responses
to the latter which, in QCD, can be expressed via the OPE to various
vacuum condensates.

The external field will contribute in two different ways: by directly
coupling to the nucleon fields which enter in the nucleon current
and by polarizing the QCD vacuum. This will introduce a susceptibility
$\chi$ which describes the response of the quark condensates to $S,$ 

\begin{align}
\langle\overline{u}u\rangle_{S} & =\langle\overline{u}u\rangle-\chi S\langle\overline{u}u\rangle\nonumber \\
\langle\overline{d}d\rangle_{S} & =\langle\overline{d}d\rangle+\chi S\langle\overline{d}d\rangle\label{eq:19}
\end{align}

where $\langle\overline{q}q\rangle\equiv\langle0\left\vert \overline{q}q\right\vert 0\rangle$

Using eq. \eqref{eq:14} one obtains

\begin{equation}
\chi\langle\overline{u}u\rangle=\frac{i}{2}\int d^{4}x\langle0\left\vert Tu_{3}(x)u_{3}(0)\right\vert 0\rangle\label{eq:20}
\end{equation}

The QCD expression for $\Pi_{1}^{q}(t)$ has been evaluated in \cite{Jin et al}

\begin{equation}
\Pi_{1}^{q}(t)=\left(C_{0}\ln(-t)+\frac{C_{1}}{t}+\frac{C_{2}}{t^{2}}+...\right)\label{eq:21}
\end{equation}

The constants $C_{i}$ are expressed in terms of the quark condensates,
the susceptibility $\chi$and the quark gluon mixed condensate $g_{s}\overline{q}\overrightarrow{\sigma}.\overrightarrow{G}q$
and its corresponding susceptibility $\chi_{m}$

\begin{align}
C_{0} & =\frac{\langle\overline{q}q\rangle}{4\pi^{2}},\nonumber \\
C_{1} & =\frac{4}{3}\chi\langle(\overline{q}q)^{2}\rangle-\frac{m_{0}^{2}}{24\pi^{2}}\langle\overline{q}q\rangle,\label{eq:22}\\
C_{2} & =(\chi+\chi_{m})m_{0}^{2}\langle\overline{q}q\rangle^{2}/6\nonumber 
\end{align}

To a good approximation $\chi_{m}=\chi$\cite{Jin et al}. With the
estimate

\begin{equation}
m_{0}^{2}=\frac{\langle g_{S}\overline{q}\overrightarrow{\sigma.}\overrightarrow{G}q\rangle_{0}}{\langle\overline{q}q\rangle}\widetilde{=}(.80\pm.20)GeV^{2}\label{eq:23}
\end{equation}

At low energies $\Pi_{1}^{q}(t)$ has double and single isolated poles
at the nucleon mass squared as well as a cut on the positive real
axis starting at $t_{th}=(M_{p}+m_{\pi})^{2}$

\begin{equation}
\Pi_{1}^{q}(t)=\frac{-2\lambda^{2}M_{p}.U}{(t-M_{p}^{2})^{2}}+\frac{A}{(t-M_{p}^{2})}+...\label{eq:24}
\end{equation}

where $\lambda$ is the coupling of the nucleon to its current

\begin{equation}
\langle0\left\vert \eta_{p}\right\vert p\rangle=\lambda U_{p}\label{eq:25}
\end{equation}

In order to relate the residue $U$ to the QCD condensates consider
the integral of the product $\Pi_{1}^{q}(t)F(t)$ over a contour $C$
in the complex $t$ plane. The contour $C$ consists of two straight
lines parallel to the cut, sandwiching it from above and below, and
running from threshold to a value $R$ and a circle of radius $R$
taken large enough to allow the replacementof $\Pi_{1}^{q}(t)$ by
its asymptotic form eq. \eqref{eq:21} on it.

$F(t)$ is a so far arbitrary entire function. Using next Cauchy's
theorem we obtain an expression relating the residue at the pole to
an integral over the cut plus an integral over the circle where the
QCD expression for the amplitude can be used. The integral over the
cut cannot be evaluated as the information on the integrand over the
continuum is scarce. We shall use the arbitrariness in the choice
of $F(t)$ to minimize this contribution so it can be safely neglected.

First take

\begin{equation}
F(t)=\left(1-\frac{t}{M_{p}^{2}}\right)P(t)\label{eq:26}
\end{equation}

in order to eliminate the contribution of the simple pole which represents
nucleon to continuum transitions. Cauchy's theorem then yields

\begin{equation}
2\lambda^{2}M_{p}UP(M_{p}^{2})=\frac{1}{\pi}\int_{th}^{R}dtF(t)Im\Pi_{1}^{q}(t)+\frac{1}{2i\pi}\oint dtF(t)(\Pi_{1}^{q}(t))_{QCD}\label{eq:27}
\end{equation}

The second integral in the expression above runs over the circle of
large radius $R.$

The coupling $\lambda$ itself can be obtained from the nucleon mass
sum rule \cite{Ioffe}\cite{CDKS}\cite{Sadovnikova} using $P(t)$
as an integration kernel with the result

\begin{equation}
-M_{p}\lambda^{2}P(M_{p}^{2})=\frac{1}{\pi}\int_{th}^{R}dtP(t)Im\Pi_{2}(t)+\frac{1}{2i\pi}\oint dtP(t)\Pi_{2}^{QCD}(t)\label{eq:28}
\end{equation}

with

\begin{equation}
\Pi_{2}^{QCD}(t)=B_{3}t\ln(-t)+\frac{B_{7}}{t}+\frac{B_{9}}{t^{2}}+...\label{eq:29}
\end{equation}

\begin{align}
B_{3} & =\frac{1}{4\pi^{2}}\langle\overline{q}q\rangle(1+\frac{3}{2}.a)\nonumber \\
B_{7} & =-\frac{1}{12}\langle\overline{q}q\rangle\langle aGG\rangle\nonumber \\
B_{9} & =4\pi^{2}\frac{136}{81}a\langle\overline{q}q\rangle^{3}\label{eq:30}
\end{align}

$a=\alpha_{S}/\pi$ is the strong coupling constant. The ratio of
eqs. \eqref{eq:27} and \eqref{eq:28} yields the residue $U$ of
interest.

Not enough experimental data is available to allow the evaluation
of the first integrals on the r.h.s. of eqs. \eqref{eq:27} and \eqref{eq:28},
only the positions of the $\frac{1}{2}^{+}$ and $\frac{1}{2}^{-}$
resonances are known and the background is impossible to model realistically.

The choice of the function $P(t)$ aims at reducing this contribution
as much as possible in order to allow its neglect. This is achieved
by minimizing the ratio$\left|P(t)/P(M_{p}^{2}\right|$or equivalently
$\left|P(t)\right|$over the resonance region. In the vast domain
of QCD sum rules the usual choice would be $P(t)=\exp(-t/M^{2})$
where the magnitude of $M$ (the Borel mass) determines the strength
of the damping of the contribution of the integrals over the continuum.
If $M$ is small the damping is good but the contribution of the unknown
terms in the QCD asymptotic expansion of the amplitudes increases
rapidly. If $M$ increases the contribution of the unknown terms decreases
but the damping worsens. An intermediate value of $M$ has to be chosen
from stability conditions which are not always met.

Because the QCD expressions for the amplitude are (except for logarithms
) infinite series in inverse powers of $t$ $,$ $C_{n}/t^{n}$ and
because the exponential can be expanded in an infinite series of powers
of $t$ of the form $\frac{1}{n!}(\frac{t}{M^{2}})^{n}$ the integral
over the circle consists of an infinite sum of terms of the form$\frac{C_{n+1}}{n!M^{2n}}$
which become important if $M$ is too small. This prompts us to choose
for $P(t)$ a polynomial of degree $N$ which involves only a finite
number of condensates ( $C_{1},C_{2},..,C$$_{N+2}$) resulting from
the integral over the circle. In order to limit the uncertainty introduced
by the condensates $N$ has to be chosen as small as possible still
large enough to provide adequate damping of the continuum. It turns
out that a second order polynomial will do the job. This will introduce
the unknown condensates $C_{3}$ and $C_{4}$ but,as we shall see,
their contribution can be estimated and turns out to be small.

The choice adopted for $P(t)$ stems from the observation of the spectrum
of the positive and negative parity nucleon resonances N(1440), N(1535),
N(1650) and N(1710). The position of these resonances imply that the
main contribution to the integrals over the nucleon continuum arises
from the interval $I:2.0GeV^{2}\precapprox t\precapprox3.0GeV^{2}$.
So let

\begin{equation}
P(t)=1-a_{1}^{\prime}t-a_{2}^{\prime}t^{2}\label{eq:31}
\end{equation}

The coefficients $a_{1}^{\prime}$ and $a_{2}^{\prime}$ are chosen
so as to practically annihilate the contribution of the continuum
on the interval $I$, e.g. they can be chosen so as to minimize the
integral$\int_{I}dtP(t)^{2}$ which insures the smallness of $\left|P(t)\right|$.
Another choice,which yields almost identical results, would be to
minimize the sum $\sum P(t)^{2}$at the resonances.

This choice reduces the relative contribution of the integrals over
the continua to only a few percent of their initial values and allows
their neglect. Numerically

\begin{align}
a_{1}' & =.807GeV^{-2},\nonumber \\
a_{2}^{\prime} & =-.160GeV^{-4}\label{eq:32}
\end{align}

The relative damping provided by P(t) at the masses of the nucleon
resonances listed above, P(resonance)/P(nucleon) is excellent. This
ratio amounts respectively to .036, -.032, -.027, .019 which is to
be compared to the corresponding values obtained from exponential
damping (at M$^{2}=1.1GeV^{2}e.g.$), .34, .26, .19, .16. 

We have thus

\begin{equation}
F(t)=1-a_{1}t-a_{2}t^{2}-a_{3}t^{3}\label{eq:33}
\end{equation}

with

\begin{align}
a_{1} & =1.947GeV^{-2},\nonumber \\
a_{2} & =-1.084GeV^{-4},\nonumber \\
a_{3} & =.184GeV^{-6}\label{eq:34}
\end{align}

If the contributions of the nucleon continua are then neglected eqs.
\eqref{eq:27} and \eqref{eq:28} become
\begin{align}
2\lambda^{2}\frac{U}{M_{p}}P(M_{p}^{2}) & =C_{0}R(1-\frac{a_{1}}{2}R-\frac{a_{2}}{3}R^{2}-\frac{a_{3}}{4}R^{3})+C_{1}-a_{1}C_{2}\label{eq:35}\\
\lambda^{2}M_{p}P(M_{p}^{2}) & =-B_{3}R^{2}(\frac{1}{2}-\frac{a_{1}^{\prime}}{3}R-\frac{a_{2}^{\prime}}{4}R^{2})-B_{7}+a_{1}^{\prime}B_{9}\label{eq:36}
\end{align}

the ratio of which yields $U.$

The choice of $R$ is dictated by stability considerations, it should
be large enough to allow the use of the QCD expressions in the integrals
over the circle of radius $R$ but not too large to invalidate the
neglect of the integral over the continuum. The optimal value of $R$
is chosen in the stability region of both equations \eqref{eq:35}
and \eqref{eq:36} Indeed, as can be seen in Fig. 1 both terms are
practically constant for $1.5GeV^{2}\precapprox R\precapprox3GeV^{2}.$

\begin{figure}
\begin{centering}
\includegraphics{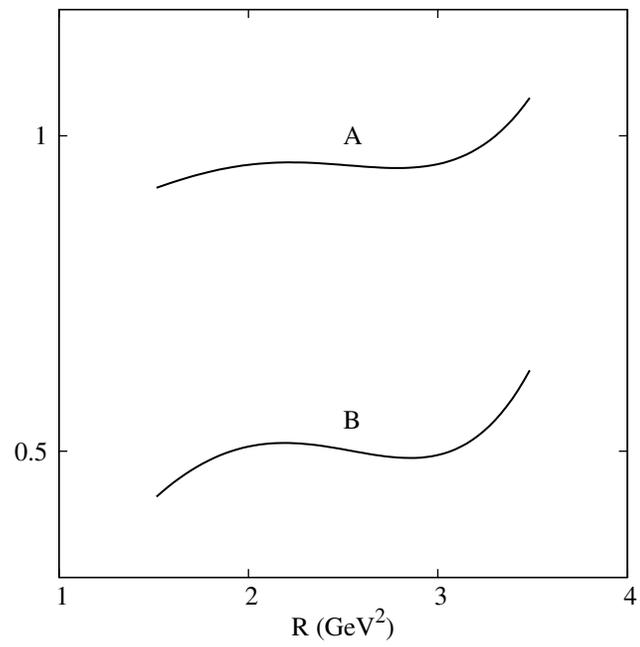}
\par\end{centering}

\caption{The right hand sides of \eqref{eq:35} (A) and of \eqref{eq:36}(B)as
a function of R (not normalized)}
\end{figure}

\section{Evaluation of the susceptibility $\chi$ }

The next task is to calculate the susceptibility $\chi$ which enters
in the expressions for the condensates. For this purpose I use a simple
model independent method \cite{cad et al}.

It follows from eq. \ref{eq:20} and from the Gell-Mann-Oakes-Renner
relation \cite{gor}

\begin{equation}
f_{\pi}^{2}m_{\pi}^{2}=-(m_{u}+m_{d})\left\langle \overline{q}q\right\rangle \label{eq:37}
\end{equation}

that

\begin{equation}
\chi=-\frac{(m_{u}+m_{d})}{f_{\pi}^{2}m_{\pi}^{2}}\psi(0)\label{eq:38}
\end{equation}

with
\begin{equation}
\psi(q^{2})=i\int d^{4}xe^{iqx}\left\langle 0\left|T(\overline{q}(x)q(x))(\overline{q}(0)q(0))\right|0\right\rangle \label{eq:39}
\end{equation}

This correlator exhibits a cut on the positive $t$ axis running from
threshold to $\infty$ and consider a contour$C$ similar to the one
used previously for the nucleon correlator. The integral over $C$
of the quantity $\psi(t)/t$ gives $\psi(0)$. As before the integration
kernel is modified so as to minimize the contribution of the $0^{+}$
continuum, i.e. consider the integral

\begin{equation}
\int_{C}dt\Delta(t)\psi(t),\label{eq:40}
\end{equation}

\begin{equation}
\Delta(t)=\frac{1}{t}-b_{0}-b_{1}t\label{eq:41}
\end{equation}

we have

\begin{equation}
\psi(0)=\frac{1}{\pi}\int_{cut}dt\Delta(t)\psi(t)+\frac{1}{2i\pi}\oint dt\Delta(t)\psi(t)\label{eq:42}
\end{equation}

In the first term on the r.h.s. of the equation above the integrand
is impossible to model realistically and mostly unknown except for
the existence of two isovector $0^{+}$ resonances at $m_{1}^{2}=.97GeV^{2}$
and $m_{2}^{2}=2.10GeV^{2}$. It is expected that the overwhelming
contribution to this integral over the cut is provided by the interval
$.90Gev^{2}\precapprox t\precapprox2.20GeV^{2}$

The choice of the constants $b_{0}$ and $b_{1}$ such that $\Delta(m_{1}^{2})=\Delta(m_{2}^{2})=0$
will practically annihilate the contribution of the integral over
the continuum so that eq. \eqref{eq:42} becomes

\begin{equation}
\psi(0)\backsimeq\frac{1}{2i\pi}\oint dt\psi_{QCD}(t)\Delta(t)\label{eq:43}
\end{equation}

The perturbative part of the correlator eq. \eqref{eq:43} is known
to 5 loops in addition to 2 non-perturbative terms\cite{chetyrkin}

\begin{equation}
\psi_{QCD}(t)=A_{0.}t\ln(-t)+\frac{A_{1}}{t}+\frac{A_{2}}{t^{2}}+...\label{eq:44}
\end{equation}

where
\[
A_{0}=-\frac{3}{8\pi^{2}}(1+\frac{11}{3}a+14.1785a^{2}+77.3535a^{3}+511.696a^{4}+...)
\]

\[
A_{1}=-\frac{1}{8}\langle aGG\rangle(1+\frac{16}{9}.a(-1GeV^{2})+\frac{121}{18}a)
\]

\begin{equation}
A_{2}=\frac{112}{27}\pi^{2}a\langle\overline{q}q\rangle^{2}\label{eq:45}
\end{equation}

So that eq. \eqref{eq:43} gives

\begin{equation}
\psi(0)=A_{0}(R-\frac{b_{0}}{2}R^{2}-\frac{b_{1}}{3}R^{3})-b_{0}A_{1}-b_{1}A_{2}\label{eq:46}
\end{equation}

In the interval $1.5GeV^{2}\precapprox R\precapprox2.5GeV^{2}.$ the
value of $\psi(0)$ oscillates between -.30 GeV and -.35 GeV which
gives an estimate of the error involved and yields

\begin{equation}
\chi=(1.03\pm.10)GeV^{-1}\label{eq:47}
\end{equation}
 Which can be compared to values appearing in the litterature which
lie in the range $.6GeV^{-1}\leq\chi\leq3GeV^{-1}$.{[}Jin et al{]}

\section{Results and Conclusions}

Before joining pieces together let us look at the values of the condensates
which enter in the $A_{i}$, $B_{i}$ and $C_{i}:$

$\langle\overline{q}q\rangle$ is obtained from the Gell-Mann, Oakes,
Renner relation\cite{gor} with the values of the quark masses $m_{u}=(2.9\pm.2)MeV$,$m_{d}=(5.3\pm.4)MeV$
\cite{CAD09}

\begin{equation}
4\pi^{2}\langle\overline{q}q\rangle=-(.79\pm.01)GeV^{3}\label{eq:48}
\end{equation}

The standard value is taken for the gluon condensate

\begin{equation}
\left\langle aGG\right\rangle =(.012\pm.006)GeV^{4}\label{eq:49}
\end{equation}

and
\begin{equation}
\langle(\overline{q}q)^{2}\rangle=\kappa\langle\overline{q}q\rangle^{2}\label{eq:50}
\end{equation}

where $\kappa$ quantifies deviations from factorization.

We have finally

\begin{equation}
U=\frac{M_{p}^{2}}{2}(C_{0}R(1-\frac{a_{1}}{2}R-\frac{a_{2}}{3}R^{2}-\frac{a_{3}}{4}R^{3})+C_{1}-a_{1}C_{2})/(-B_{3}R^{2}(\frac{1}{2}-\frac{a_{1}^{\prime}}{3}R-\frac{a_{2}^{\prime}}{4}R^{2})-B_{7}-a_{1}^{\prime}B_{9})\label{eq:51}
\end{equation}

Let us look at the sources of uncertainty in the expression above.

The first stems from the incomplete knowledge of the OPE: with our
choice of polynomials the numerator of eq. \eqref{eq:51}should be
augmented by two additional terms $a_{2}C_{3}+a_{3}C_{4}$ and the
denominator by $a_{2}^{\prime}B_{11}$. The higher order condensates
$C_{3},C_{4}$ and $B_{11}$are of course unknown but it is possible
to estimate them using the method of Pade' approximants. It turns
out that the error introduced by the neglect of the higher order condensates
amounts to no more than $2\%.$

Another source of uncertainty stems from the choice of the coefficients
$a_{1}^{\prime},\, a_{2}^{\prime}$ (and consequently of the coefficients
$a_{1},\, a_{2},\, a_{3}$ ),varying these within reasonable limit
in order to deplete the contribution of the interval $2.0GeV^{2}\precapprox t\precapprox3.0GeV^{2}$
introduces an uncertainty of $\thicksim6-7\%$ in the value of $U.$

An additional uncertainty of course is the one arising from the values
of the condensates themselves.

All these uncertainties however are overwhelmed by the one coming
from the undeterminacy in the value of $\kappa$ which measures the
deviation from factorization in the value of the four quark condensate.
There is no consensus on the value of this quantity in the litterature.
Phenomenological estimates \cite{cata} place it in the range $1\precapprox\kappa\precapprox4.$

With a central value \cite{CAD09} $\delta m=2.4MeV$, $\delta M_{q}$
is plotted against $\kappa$ in Fig. 2 where the experimental limits
are also shown.

Taking in consideration the uncertainty in $\delta m$ we see that
the experimental value of $\delta M_{q}$is reproduced for

\begin{equation}
1.7\lesssim\kappa\lesssim2.3\label{eq:52}
\end{equation}

\begin{figure}
\begin{centering}
\includegraphics{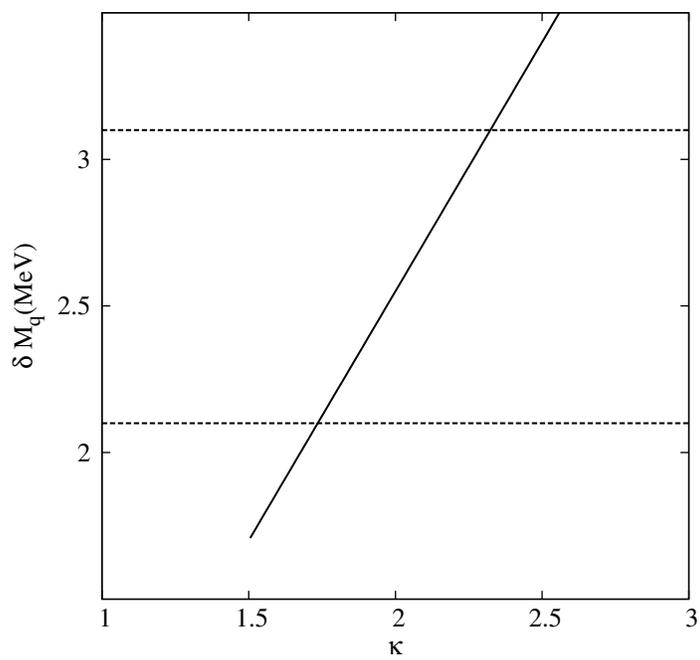}
\par\end{centering}

\caption{The strong part of the neutron-proton mass difference $\delta M_{q}$
as a function of $\kappa$}
\end{figure}

It is instructive to compare the results obtained to the ones given
by use of the exponential damping kernel $P(t)=e^{-\frac{t}{M^{2}}}$.
Expression \eqref{eq:51} for $U$ is then replaced by

\begin{equation}
U=\frac{M_{p}^{2}}{2}(C_{0}M^{4}\int_{0}^{R/M^{2}}dxe^{-x}(x-\frac{M_{p}^{2}}{M^{2}})-M_{p}^{2}C_{1}+(1+\frac{M_{p}^{2}}{M^{2}})C_{2})/(B_{3}M^{4}\int_{0}^{R/M^{2}}dxxe^{-x}+B_{7}-\frac{B_{9}}{M^{2}})\label{eq:53}
\end{equation}

which reproduces qualitatively the same results as before but which
fails to show good stability in variations of the Borel mass $M^{2}$
as can be seen from Fig. 3. This instability stems mostly from the
fact that the contribution of the nucleon continuum remains important.
In \cite{Jin et al} the contribution of the single pole in eq.\eqref{eq:24}
is not eliminated but taken into account. This introduces an additional
parameter and the accuracy of the method used to estimate it is hard
to assess

\begin{figure}
\begin{centering}
\includegraphics{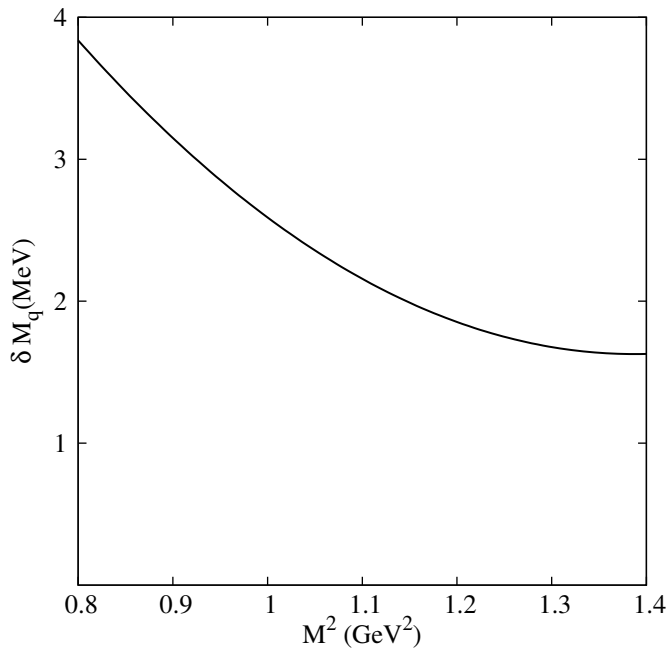}
\par\end{centering}

\caption{$\delta M_{q}$ as a function of the Borel mass $M^{2}$for $\kappa=2$}
\end{figure}

To conclude, I have presented an evaluation of the neutron-proton
mass difference $\delta M_{np}$ using polynomial kernels in dispersion
integrals taylored to reduce the contributions of the unknown parts
of the continua to an extent which allows their neglect and which
guarantees the stability of the calculation.

The numerical result depends sensitively on the value of $\kappa$
which quantifies the deviation of the value of the four quark condensate
from the one given by the factorization assumption and reproduces
the experimental value of $\delta M_{np}$ for$\kappa\sim2.$ The
old value $\delta M_{coul}=-(.76$$\pm.30)MeV$ would correspond to
values of $\kappa$ closer to unity.

It is finally worth to note that a recent analysis of Weinberg type
sum rules \cite{hohler} yields $\kappa=2.1_{-.2}^{+.3}$

\pagebreak{}

\end{document}